# Spatial Imaging of Landé g Factor through Extended Kondo Effect in a Single Magnetic Molecule

Liwei Liu,[1] Yuhang Jiang,[1] Boqun Song,[1] Kai Yang,[1] Wende Xiao,[1] Shixuan Du,[1] Min Ouyang,[2] Werner A. Hofer,[3] Antonio. H. Castro Neto[4], and Hong-Jun Gao[1*]

[1] *Institute of Physics, Chinese Academy of Sciences, P.O. Box 603, Beijing 100190, China*

[2] *Department of Physics and Center for Nanophysics and Advanced Materials, University of Maryland, College Park, MD 20742, USA*

[3] *Department of Physics, The University of Liverpool, Liverpool L69 3BX, UK*

[4] *Graphene Research Centre, Department of Physics, National University of Singapore, 117542, Singapore*

A methodology of atomically resolved Landé g factor mapping of a single molecule is reported. Mn(II)-phthalocyanine (MnPc) molecules on Au(111) surface can be dehydrogenated via atomic manipulation with manifestation of tailored extended Kondo effect, which can allow atomically resolved imaging of the Landé g factor inside a molecule for the first time. By employing dehydrogenated MnPc molecules with removal of six H atoms (-6H-MnPc) as an example, Landé g factor of atomic constituents of the molecule can be obtained, therefore offering a unique g factor mapping of single molecule. Our results open up a new avenue to access local spin texture of a single molecule.

The magnetic property of a nanostructure plays pivotal role in the design of miniaturized spintronic devices [1]. The Kondo resonance, arising from many-body spin-flip scattering between a local magnetic moment and the conduction electrons of a host metal, has driven much interest in recent decades [2, 3]. One signature of a Kondo resonance is the splitting of its peak position at the presence of external magnetic field, by doubling the ordinary Zeeman splitting [4]. Such magnetic field induced Kondo splitting has been demonstrated to enable direct measurement of Landé g factor in quantum dots in both transport and scanning tunneling microscope (STM) measurements [5-9]. This, in principle can allow understanding of internal spin texture in a complex molecule, if the Kondo splitting can be finely resolved inside a molecule with atomic details.

In this Letter, directly enabled by the STM's capability to visualize and modify molecule down to the atomic level, we have systematically investigated the Kondo effect in a series of dehydrogenated (DH-) MnPc molecules, and observed that Kondo effect can be spatially extended to non-magnetic atom constituents of the molecule in addition to the central $Mn^{2+}$ ion. This effect can be explained by charge transfer induced spin polarization that agrees well with Density Functional Theory (DFT) calculation. As a result, by measuring the Kondo splitting under an applied magnetic field, Landé g factors of individual atoms within a single molecule can be exclusively achieved and show pronounced variations for manganese, carbon and nitrogen atoms. Because Landé g factor is relevant to local chemical environment, our result can provide a valuable avenue to reveal internal molecular structure, which otherwise is challenging to achieve.

In the experiments we deposited MnPc molecules on the Au(111) substrate at room temperature, and subsequently cooled the sample down to 4.2 K for STM measurements. The molecular structure of MnPc is shown in the left panel of Fig. 1(a). On a terrace, each MnPc molecule appears as a four-lobed shape with a central protrusion (see the right panel of Fig. 1(a), and Fig. S1(a)), which is consistent with the molecular $D_{4h}$ symmetry and indicates a flat-lying adsorption configuration.

Molecule surgery enabled by STM based manipulation represents an efficient way to alter the molecular structure and corresponding electronic properties [10-12]. We modify the MnPc molecule by selectively removing its hydrogen atoms, which can be achieved by placing the STM tip above the specific lobe (as indicated by the red dotted circle in the left panel of Fig. 1(a)) and applying a voltage pulse of 3.6 V for 1 second with feedback off (STM condition: 0.1 V/0.1 nA). Fig. 1(b) shows consecutive topographic images of dehydrogenated MnPc with well-regulated H atoms (also see Fig. S1(b)). We further define the modified molecules as -2H-MnPc, -4H-MnPc, -6H-MnPc, -8H-MnPc, corresponding to molecules after removing two, four, six and eight hydrogen atoms, respectively.

In conjunction with topography change of the dehydrogenated molecules, *dI/dV* spectra (Fig. 1(c)) acquired at the central $Mn^{2+}$ ion also show intriguing evolution of the local density of states (LDOS) near the Fermi level, from a step-like feature for intact MnPc (the bottom curve in Fig. 1(c)) to a sharp peak for -8H-MnPc (the top curve in Fig. 1(c)). We can attribute this characteristics to the Kondo resonance, which arises from spin-flip scattering between the local magnetic impurities, the $Mn^{2+}$ ions, and the Fermi sea of electrons at the substrate [13-18]. One of the hallmarks of a Kondo resonance is the splitting of its peak position at the presence of external magnetic field B [4-9]

$$\Delta E = 2g\mu_B B \quad (1)$$

where $\mu_B = 58 \mu eV/T$ is the Bohr magneton and g is the Landé g factor. We measure the *dI/dV* spectra under applied magnetic fields, and indeed observe the splitting (see Figs. S2 and S3), which confirm the Kondo resonance mechanism.

To quantitatively analyze the evolution of the Kondo resonances with dehydrogenation, we fitted experimental data with a Fano function [19]

$$\frac{dI}{dV}(V) = A \cdot \frac{(\varepsilon+q)^2}{1+\varepsilon^2} + B \quad (2)$$

where $\varepsilon$ is the normalized energy, $\varepsilon = (eV - \varepsilon_0)/\Gamma$ ($\varepsilon_0$ is the position of the resonance and $\Gamma$ is the half-width at half-maximum of the Kondo peak), $T_K$ is the Kondo temperature ($k_B T_K = \Gamma$) and *q* is the Fano factor, which is given by

$$q = \frac{t_2}{\pi \rho_0 V t_1} \quad (3)$$

Here, $t_1$ and $t_2$ are the matrix elements for electron tunneling into continuum bulk states and the discrete Kondo resonance, respectively [19].

For the intact MnPc, an asymmetric Kondo resonance is detected, yielding a negative *q* of -1.08±0.1 and $T_K$ of 64±3 K (Fig. 1(d)). As the number of removed peripheral H atoms increases from two to six, the Kondo temperature $T_K$ decreases from 64 K to 30 K, but increases to 39 K when eight H atoms are removed. The variation of shape factor *q* with the number of removed H atoms follows the similar trend of that of Kondo temperature $T_K$.

To rationalize experimental results we have performed the DFT calculations on a series of dehydrogenated MnPc molecules. The DFT calculations show that the distance between the $Mn^{2+}$ ion and the substrate increases from 3.2 Å for intact MnPc to 5.2 Å for -6H-MnPc and then decreases to 4.9 Å for -8H-MnPc (Fig. 1(e)). Because the magnetic moment of the $Mn^{2+}$ ion varies only little in the process of dehydrogenation, the change of the distance between the $Mn^{2+}$ ions and the substrate determines their coupling strength: a larger distance results in a weaker interaction between the $Mn^{2+}$ ions and the Au substrate, and hence a reduced Kondo temperature [20]. Our DFT calculation shows good agreement with experimental observation of $T_K$ and $q$.

To gain insight into the spatial distribution of the molecular Kondo effect, we perform the *dI/dV* mapping over both the intact and the dehydrogenated MnPc molecules near the Fermi level. The *dI/dV* image of an intact MnPc (Fig. S4(a)) shows a bright spot at the molecule's center, suggesting the resonance is localized at the $Mn^{2+}$ ion of the molecule. The local Kondo effect on a magnetic ion in our case is also consistent with previous work [15, 16]. Similar feature is also observed for the -2H and -8H-MnPc molecules (Figs. S4(b) and S4(c)).

Importantly, the *dI/dV* map of -4H-MnPc (Fig. 2(a)) shows the appearance of two additional bright spots on the remaining lobes other than the molecular center, with evident peak feature in the corresponding *dI/dV* spectrum (Fig. 2(e). Similar observation is also made with -6H-MnPc molecule (Fig. 2(b) and 2(e)). Intuitively, observed extended Kondo effect suggests the spin-polarization at the lobe, which may arise from the charge transfer within the molecule after dehydrogenation [21]. To check the validity of this mechanism, we have performed DFT calculations with the -4H-MnPc and -6H-MnPc molecules. The projected spin density of states for -4H-MnPc indeed shows that the spin polarization can be found not only at the $Mn^{2+}$ ion but also at the benzene ring on the remaining two lobes (see Fig. 2(c)), thus Kondo resonance can be expected at these sites because it arises from the interference between the local spin and the conduction electron of substrate. The DFT results of the -6H-MnPc molecules also show spin polarization at the remaining lobe, consistent with our experimental observation that the -6H-MnPc molecule shows extended Kondo resonance at the lobe (see Fig. 2(d)).

The attribution of this spatially extended feature to the Kondo effect can be confirmed by the splitting of the *dI/dV* peaks under external magnetic fields (Fig. 3), which also excludes other possibilities including such as molecular orbitals [22, 23]. The measured energy splitting has a linear relationship with the applied magnetic field, and the linear fitting can uniquely determine corresponding g factors (see Fig. 3). We observe the g factors of 2.07 and 2.17 for the center (corresponding to $Mn^{2+}$ ion) and the lobe (C atoms), respectively. Similar magnetic field dependent measurement performed on different nitrogen atoms within -6H-MnPc molecule (labeled as N1, N2 and N3 for clarity) also leads to corresponding g factors as 2.12, 2.41 and 1.81, respectively. These g-factors are all different from that of free electrons, which represent local chemical environment inside a molecule.

In summary, while the Landé g factor of molecules and nanoscale structures have been studied extensively over the past decades, we have demonstrated for the first time measurement of g-factor of atomic constituents inside a single molecule, thus offering a unique way to probe local spin texture (such as local spin-orbit interaction) within the molecule. This is achieved by an extended Kondo effect through the charge transfer induced spin polarization mechanism, which can be very general to many magnetic molecular structures. The ability to access local spin texture inside a molecule provides not only fundamental insight of spin-orbit interactions with atomic resolution but also new methodology to manipulate and design functionality of molecules.

**Figure Captions**

FIG. 1. Manipulation of molecular Kondo effect by systematic dehydrogenation. (a) Molecular structure (left) and STM images of intact MnPc molecules. Scanning bias: -0.2 V; Tunneling current: 10 pA; Scale bar, 1 nm. (b) STM images of dehydrogenated MnPc molecules. The two outer hydrogen atoms of each of the four lobes (as indicated by the dotted circle in (a)) were removed step by step. Scanning bias: -0.2 V; Tunneling current: 8.6 pA; Scale bar, 1 nm. (c) Corresponding dI/dV spectra of the molecules in (a). The spectra were taken at the position of $Mn^{2+}$ ions. Each successive data is vertically shifted 0.2nA/V for clarity. The black curves are Fano-fittings of the Kondo resonance. (d) The evolution of Kondo temperature, shape factor of molecules derived from Fano fitting of dI/dV spectra. (e) Calculated distance between $Mn^{2+}$ ion and Au substrate (the first layer) with increasing number of removed H atoms.

FIG. 2. *dI/dV* mappings of intact and dehydrogenated MnPc molecules. (a) and (b) *dI/dV* mappings of -4H and -6H-MnPc molecules. *dI/dV* mappings were taken at 6 mV; Scale bar, 1 nm. The dotted lines are molecular contours. (c) and (d) Calculated spin polarization of -4H-MnPc and -6H-MnPc, overlayed on the *dI/dV* mappings, showing the extended projected Kondo effect. (e) *dI/dV* spectra of -4H and -6H-MnPc molecules. The spectra are got at the lobe indicated by the black cross in (a) and (b), respectively. Each successive data is vertically shifted 0.2 nA/V for clarity. The black

curves are Fano-fittings of the Kondo resonance.

FIG. 3. g factor mapping at a -6H MnPc molecule. (a) Topography of a -6H-MnPc molecule and overlay of corresponding model. Scale bar, 0.5 nm. (b) to (f) *dI/dV* spectra got on different sites corresponding to the marked positions of the -6H-MnPc molecule in (a). All *dI/dV* spectra were measured at the sample temperature of 0.4 K under a magnetic field of Bz= 0～11 T. The Kondo peaks split at higher magnetic fields. Each successive data is vertically shifted with 0.2 nA/V for clarity. (g) g factors calculated from the Kondo splitting from (a) to (e) (for the linear fit, see Fig. S4).

## Acknowledgements

Work at IOP was supported by grants from National Science Foundation of China, National "973" projects of China, the Chinese Academy of Sciences, and SSC. Work in Liverpool was supported by the EPSRC Car-Parinello consortium, grant No EP/F037783/1. M.O. acknowledged support from the ONR (award #: N000141110080). AHCN acknowledges DOE grant DE-FG02-08ER46512, ONR grant MURI N00014-09-1-1063, and the NRF-CRP award "Novel 2D materials with tailored properties: beyond graphene" (R-144-000-295-281).

# Figures

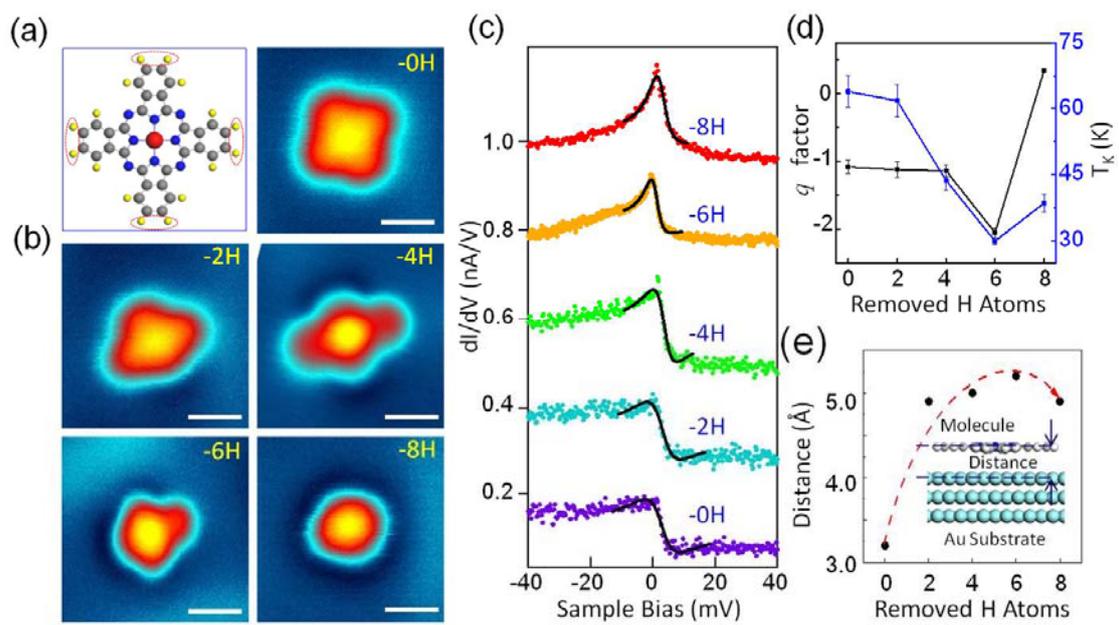

**FIG. 1**

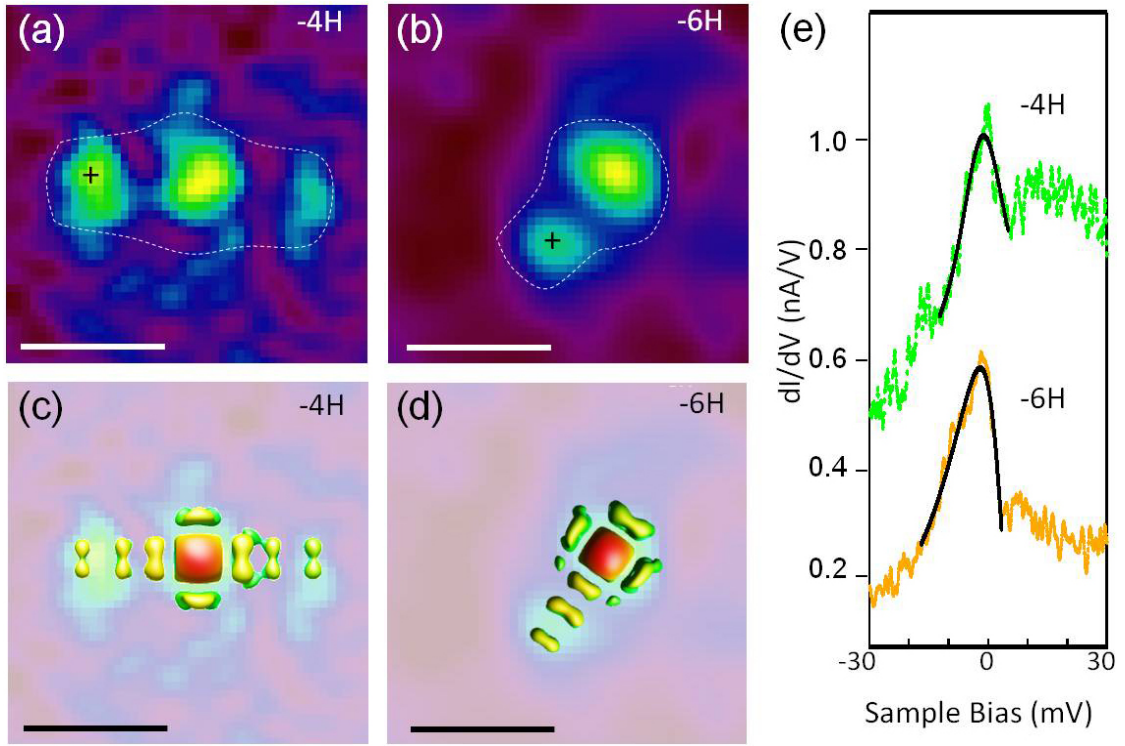

**FIG. 2**

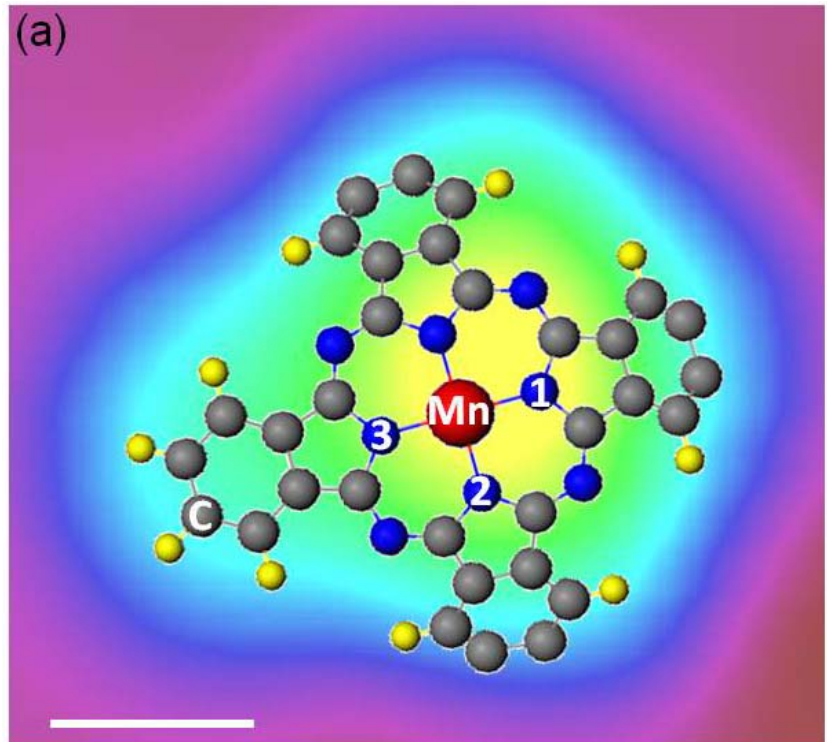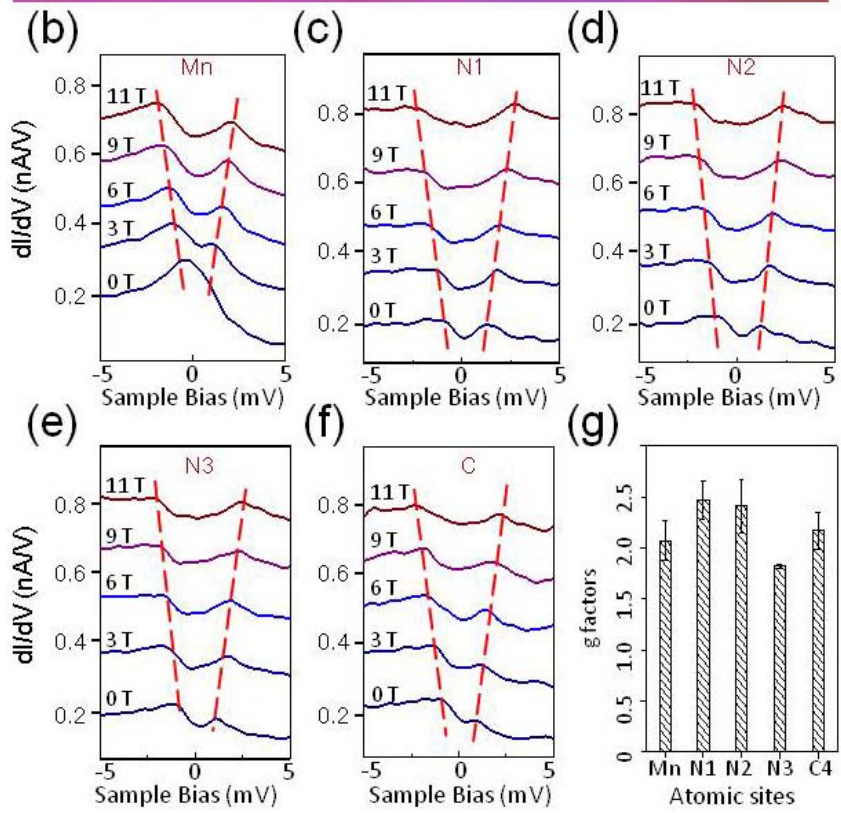

**FIG. 3**

Supplemental Material

# Spatial Imaging of Landé g Factor Through Extended Kondo Effect in a Single Magnetic Molecule

Liwei Liu,[1] Yuhang Jiang,[1] Boqun Song,[1] Kai Yang,[1] Wende Xiao,[1] Shixuan Du,[1] Min Ouyang,[2] Werner A. Hofer,[3] Antonio. H. Castro Neto[4], and Hong-Jun Gao[1*]

[1] *Institute of Physics, Chinese Academy of Sciences, P.O. Box 603, Beijing 100190, China*

[2] *Department of Physics and Center for Nanophysics and Advanced Materials, University of Maryland, College Park, MD 20742, USA*

[3] *Department of Physics, The University of Liverpool, Liverpool L69 3BX, UK*

[4] *Graphene Research Centre, Department of Physics, National University of Singapore, 117542, Singapore*

\* Email: hjgao@iphy.ac.cn.

## I. Experimental methods

The experiments were conducted with an ultrahigh vacuum (UHV) scanning tunneling microscopy (STM) system (Unisoku) at a base temperature of 0.4 K by means of a single-shot $^3$He cryostat. A magnetic field up to 11 T can be applied perpendicular to the sample surface. By an elaborate grounding and shielding scheme, the stability of the tunneling junction is greatly enhanced and a high spectroscopic resolution of ~0.1 meV is achieved at 0.4 K. Clean Au(111) surfaces were prepared by repeated cycles of argon sputtering and annealing to 580 K for 10 minutes. After purification, MnPc molecules were thermally evaporated onto a Au(111) surface at room temperature. All STM images were acquired in constant-current mode with etched W tips, and the given voltages refer to the sample. Scanning tunneling spectroscopy (STS) and conductance (*dI/dV*) maps were measured using a lock-in amplifier with a bias modulation of 0.5 mV at 973 Hz.

## II. Density Functional Theory

DFT calculations were performed by Vienna ab-initio simulation package (VASP) [1, 2] Perdew-Wang 91 exchange correlation functional and projector augmented-wave method were employed, and the energy cutoff for plane-wave basis set was 400 eV. A *c*(7×8) supercell containing three layers of gold atoms was employed to model an isolated molecule on Au(111) substrate. To ensure that interactions between the periodic slabs through vacuum are negligible, the slabs were separated by a vacuum gap of 20 Å. For geometry optimization purposes, the bottom layer was fixed, while the adsorbate and the other metal layers were allowed to relax until the forces were less than 0.01 eV/Å on each atom. A single Γ point was used in sampling the Brillouin zone due to the numerical limitations.

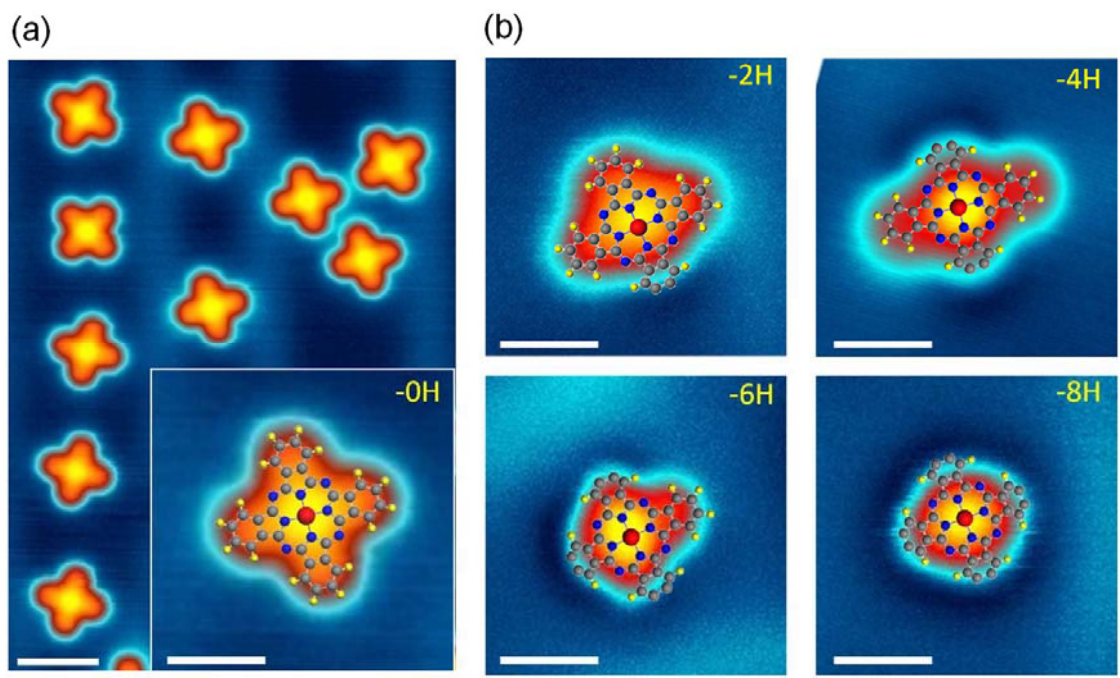

FIG. S1. STM images of intact and dehydrogenated MnPc molecules on Au(111).(a) STM images of MnPc molecules on Au(111). Scanning bias: -3 V; Tunneling current: 7 pA; Scale bar, 2 nm. The inset is an overlay of a MnPc molecular model on top of the STM image of intact MnPc molecule, and the scale bar is 1 nm. (b) STM images of dehydrogenated MnPc molecules and overlay of corresponding model. Scanning bias: -0.2 V; Tunneling current: 8.6 pA; scale bar, 1 nm.

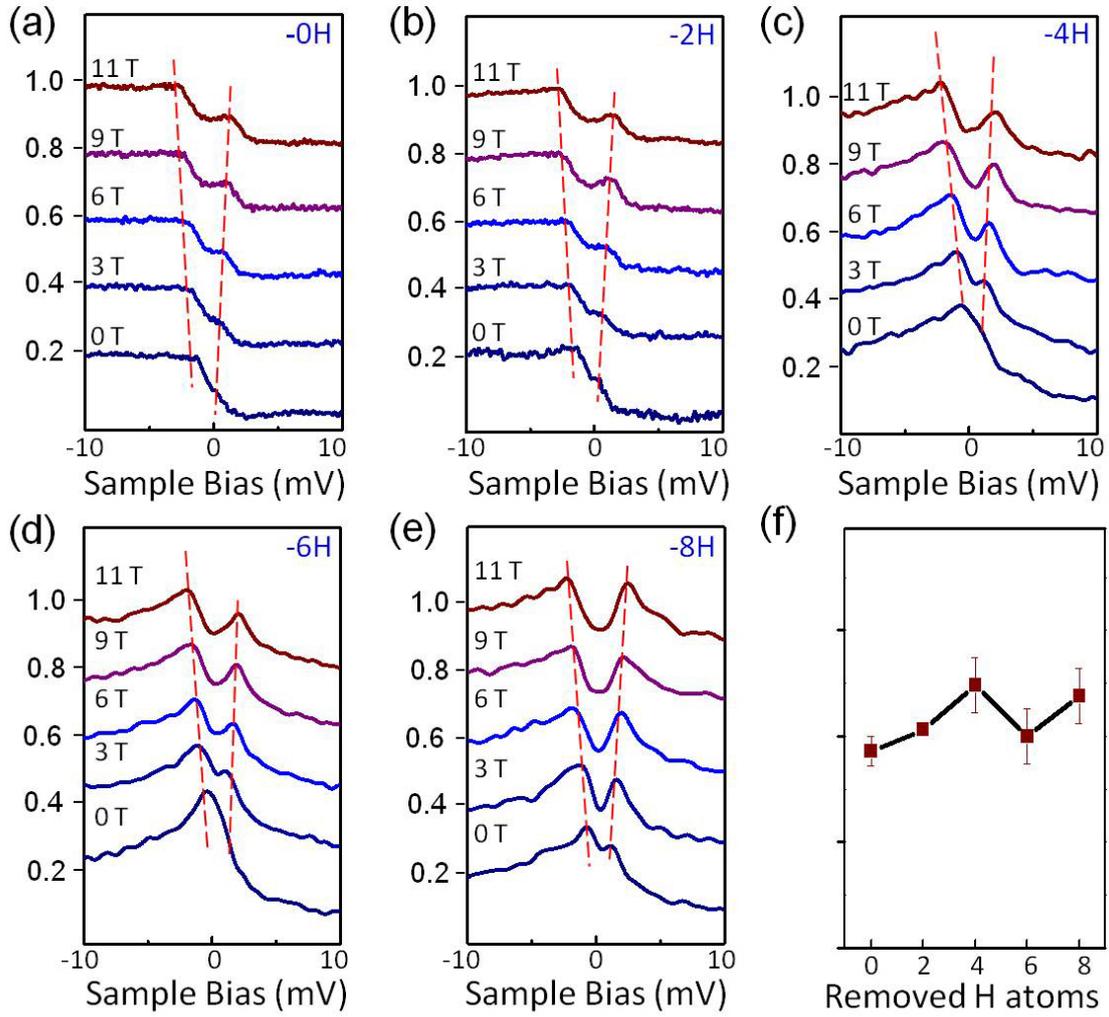

FIG. S2. Splitting of the Kondo resonance with applied magnetic field up to 11 T. (a) to (e) *dI/dV* spectra with tip positioned at the $Mn^{2+}$ ion of the intact and dehydrogenated MnPc molecules. All *dI/dV* spectra were measured at the sample temperature of 0.4 K under a magnetic field of $B_z$= 0～11 T. The Kondo peak splits asymmetrically at higher magnetic fields. Note that for intact, -2H and -4H-MnPc molecules, zero field splitting is observed. (f) g factors calculated from the Kondo splitting in (a) to (e) (for the linear fit, see Fig. S4).

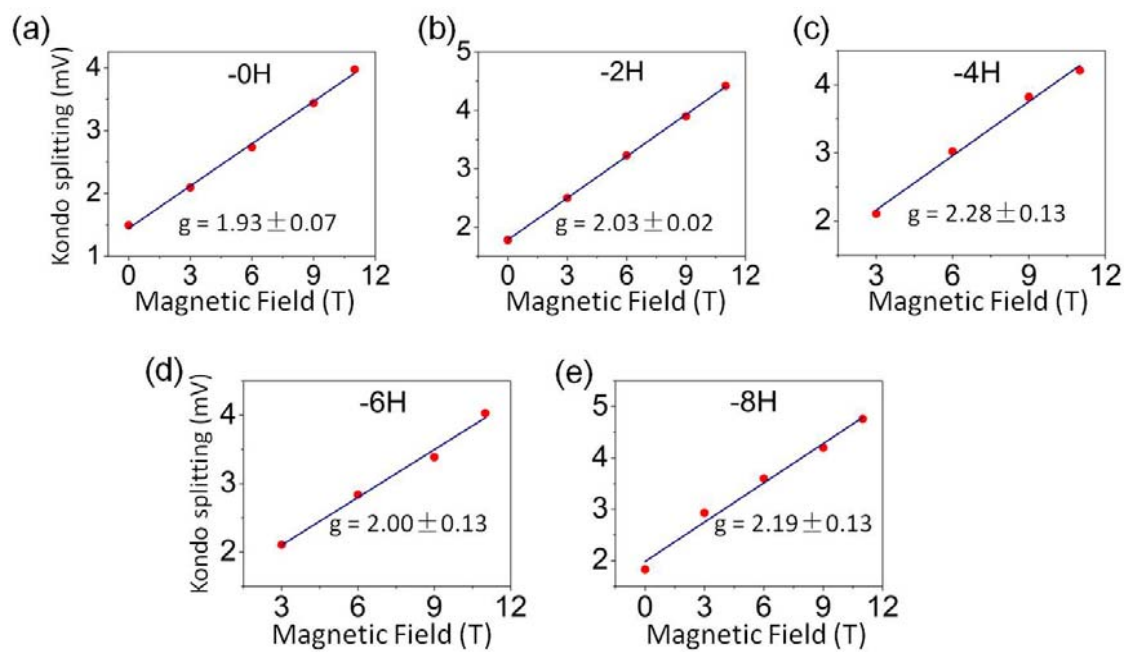

FIG. S3. Magnetic field dependence of the Kondo splitting at $Mn^{2+}$ ion for intact and DH-MnPc molecules, showing the fitting of g factors.

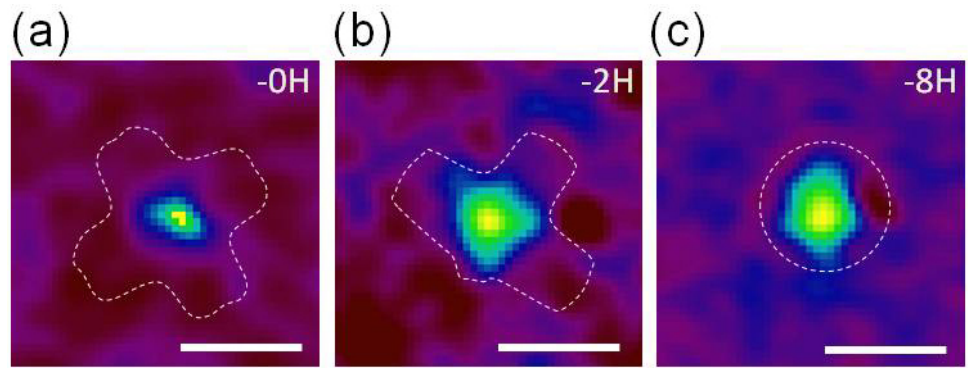

FIG. S4. *dI/dV* mappings of intact and dehydrogenated MnPc molecules. (a) to (c) *dI/dV* mappings of intact, -2H, and -8H-MnPc molecules. *dI/dV* mappings were taken at 6 mV; Scale bar, 1 nm. The dotted lines in upper panels are molecular contours.

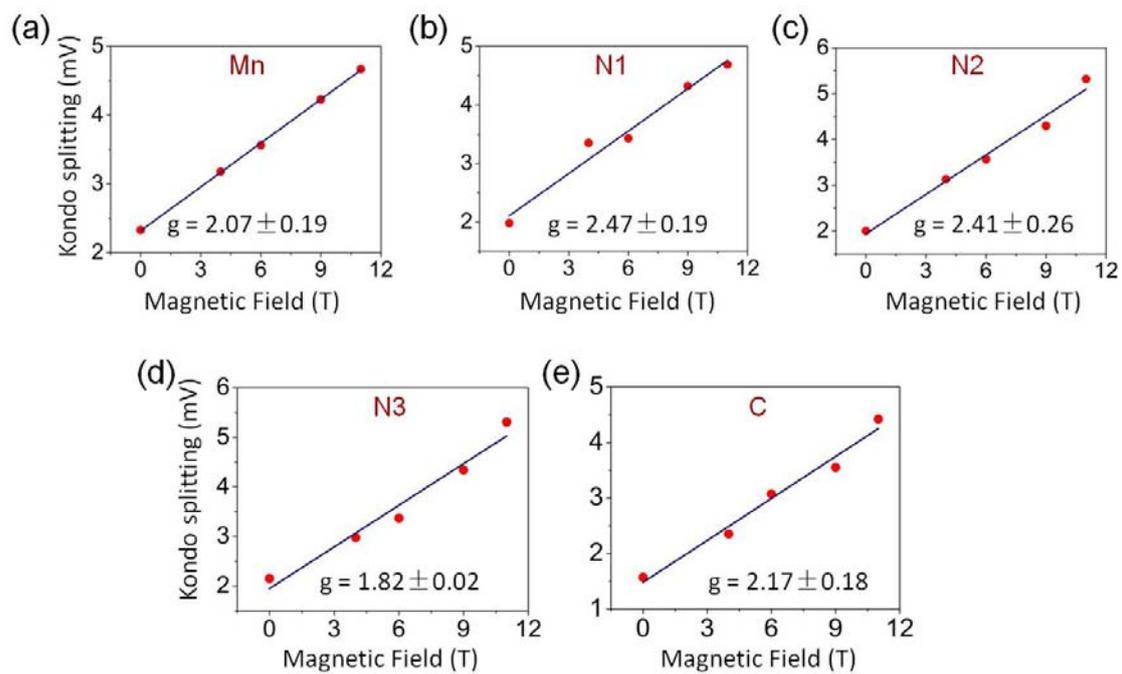

FIG. S5. Magnetic field dependence of the Kondo splitting at different positions in a -6H-MnPc molecule in Fig. 3, showing the fitting of g factors.